\documentclass{jfm}

\usepackage{graphicx}
\usepackage{natbib}
\usepackage[dvips]{color}

\ifCUPmtlplainloaded \else
  \checkfont{eurm10}
  \iffontfound
    \IfFileExists{upmath.sty}
      {\typeout{^^JFound AMS Euler Roman fonts on the system,
                   using the 'upmath' package.^^J}%
       \usepackage{upmath}}
      {\typeout{^^JFound AMS Euler Roman fonts on the system, but you
                   dont seem to have the}%
       \typeout{'upmath' package installed. JFM.cls can take advantage
                 of these fonts,^^Jif you use 'upmath' package.^^J}%
      }
  \else
  \fi
\fi

\ifCUPmtlplainloaded \else
  \checkfont{msam10}
  \iffontfound
    \IfFileExists{amssymb.sty}
      {\typeout{^^JFound AMS Symbol fonts on the system, using the
                'amssymb' package.^^J}%
       \usepackage{amssymb}%

      }{}
  \fi
\fi

\ifCUPmtlplainloaded \else
  \IfFileExists{amsbsy.sty}
    {\typeout{^^JFound the 'amsbsy' package on the system, using it.^^J}%
     \usepackage{amsbsy}}
    {\providecommand\boldsymbol[1]{\mbox{\boldmath $##1$}}}
\fi

\newcommand\Ret{\mbox{\textit{Re}$_\tau$}}  % Reynolds number
\newcommand\Rem{\mbox{\textit{Re}$_{\rm m}$}}  % Reynolds number
\newcommand\Rec{\mbox{\textit{Re}$_{\rm c}$}}  % Reynolds number
\newcommand\Prn{\mbox{\textit{Pr}}}  % Prandtl number, cf TeX's \Pr product
\newsavebox{\astrutbox}
\sbox{\astrutbox}{\rule[-5pt]{0pt}{20pt}}

\newcommand{\fref}[1]{figure~\ref{#1}}

\newcommand{\tref}[1]{table~\ref{#1}}

\title[Channel flow at transitional Reynolds numbers]{Turbulent heat transfer in a channel flow \\at transitional Reynolds numbers}

\author[T. Tsukahara, H. Kawamura]{
Takahiro Tsukahara and Hiroshi Kawamura\thanks{Present affiliation: Tokyo University of Science, Suwa}}

\affiliation{Department of Mechanical Engineering, Tokyo University of Science, \\Yamazaki 2641, Noda-shi, Chiba, 278-8510, Japan}

\date{}

\begin{document}

\maketitle

\begin{abstract}

Direct numerical simulation of a turbulent channel flow with heat transfer was performed at very low Reynolds numbers. Two different thermal boundary conditions were studied, and temperature was considered as a passive scalar. The computations were carried out with huge computational boxes (up to $327.7 \times 2 \times 128$ in the streamwise, wall-normal, and spanwise directions, respectively). The emphases of this paper are to investigate the large-scale structure (puff) in the intermittent-turbulent flow including the scalar fields and to provide the values of the transitional and critical Reynolds numbers, below which the turbulent flow becomes intermittent and laminar, respectively. The statistics, such as the skin friction and the Stanton number, were also examined: they suggest that the puff should be effective in sustaining turbulence and in heat transfer enhancement.

\end{abstract}

\section{Introduction}

Turbulent heat transfer in a transitional channel flow occurs in a variety of heat-exchange processes such as coolant flows in a high temperature gas-cooled nuclear reactor or in a micro-scale heat exchanger. Often in such applications, low-Reynolds-number flows are employed to obtain a high outlet temperature while it maintains turbulence for effective heat transfer. As is well known, most channel flows in various engineering systems can undergo laminar-to-turbulent transition below the critical Reynolds number given by the linear instability analysis. Also, in the fundamental flow physics, the process of laminarization is important. Hence, the subcritical transition from turbulence to laminar has been studied experimentally by a number of researchers.

We often observe that fluid flows at transitional Reynolds numbers are `locally' turbulent, which implies that fluid regions in laminar and turbulent states co-exist close to each other. This type of transition gives rise to a turbulent equilibrium puff, which preserves itself and occupies an entire cross-section in the case of the pipe flow \citep[cf.,][]{Wygnanski73,Wygnanski75}. 
Recently, a similar structure of the puff was found in the channel flow obtained by direct numerical simulation (DNS). 
\cite{Tsukahara2005, Tsukahara2006} demonstrated that there occurred an isolated-turbulent structure (puff) at $\Ret = 80$ with the use of a large computational box of $51.2\delta \times 2\delta \times 22.5\delta$, as we were able to show the existence of the puff. Here, the Reynolds number of $\Ret$ is based on the friction velocity and the channel half width ($\delta$). The ensemble-averaged structure of the puff has been studied in the previous paper \citep{Tsukahara2006}, in which the computational box captured one wavelength of the puff in both stream- and spanwise directions. However, we must apply a further enlarged domain to allow the periodic boundary condition. As presented in this paper, we employed the large enough box size compared to a prospective size of the puff.

During the past twenty years, DNS of turbulent heat transfer in the channel flow has advanced from low-Reynolds-number simulations at low/moderate Prandtl numbers (i.e., for $\Ret = 150$--180 and $\Prn = 0.025$--5.0 \citep{Kim89, Lyons91, Kasagi92, Kawamura98}) to simulations at high Reynolds- and/or Prandtl numbers \citep{Na00, Kawamura00, Abe04a, Abe04b}. However, almost none have been accomplished for the heat transfer at transitional Reynolds numbers with the intermittent structure, so-called puff, as applied with a large computational box.

In the present work, the various statistics associated with fully developed scalar fields for two different thermal boundary conditions are presented and discussed with emphasis on the puff. A series of DNS has been made for $\Ret = 56$--150 with the large computational box sizes, as summarized in \tref{tab:hugebox}, to capture the large-scale structure of the puff.

\section{Numerical procedure}

%%-------------------------------------- 
\begin{figure}
 \begin{center}
  \centerline{\includegraphics[width=110mm]{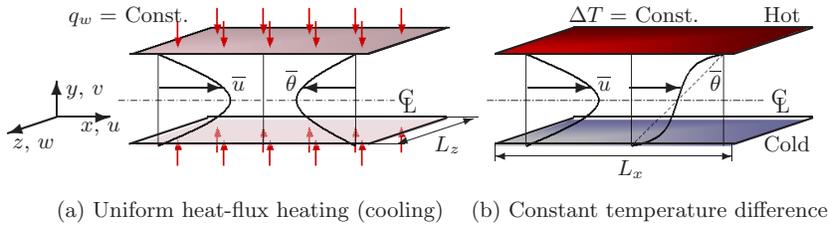}}
  \caption{Configuration of DNS.}
  \label{fig:config}
 \end{center}
\end{figure}

The configuration is the fully developed channel flow, which is driven by a uniform pressure gradient. As shown in \fref{fig:config}, one of the thermal boundary conditions is the uniform heat-flux heating over the both surfaces (hereafter, UHF), and the other one is the constant temperature difference between the walls (CTD). Note that the streamwise, wall-normal, and spanwise locations (velocities) are described by $x$ ($u$), $y$ ($v$), and $z$ ($w$), respectively. The periodic boundary condition is imposed in the $x$ and $z$ directions. The non-slip condition is applied on the walls. Uniform grid mesh is used in the horizontal directions, and non-uniform mesh in the y direction. For the air of Prandtl number $\Prn = 0.71$, all fluid properties are treated as constant. For the spatial discretization, the finite difference method is adopted. The fundamental equations are the continuity, Navier-Stokes and energy equations. For the energy equation in the case of UHF, the temperature difference $\Delta T$ is introduced to be satisfied with the constant heat-flux boundary condition. Since the statistically averaged temperature increases linearly in the streamwise direction, the instantaneous temperature $T$ can be divided into the mean part and the fluctuation one, as follows:
\begin{equation}
T(x,y,z)=\frac{{\rm d} T_{\rm w}}{{\rm d} x} - \theta (x,y,z)
\end{equation}
where $T_{\rm w}$ is the wall temperature. Further details of the numerical scheme can be found in \cite{Kawamura00}. 
The computational conditions, such as the box size, the grids number and the resolution, are given in tables \ref{tab:hugebox} and \ref{tab:middlebox}. Simulations were performed with `huge' computational box sizes at $\Ret$ ranging from 150 down to 56 (see \tref{tab:hugebox} for overview of the simulations), while six simulations at $\Ret = 64$--180 described in \tref{tab:middlebox}, see also \cite{Tsukahara2006} were performed with usual box sizes (`medium' box). The obtained Reynolds numbers of $\Rem$, based on the bulk mean velocity ($u_m$) and $2\delta$, and $\Rec$, based on the channel centerline velocity ($u_c$) and $\delta$, are also shown in the tables. As for the initial velocity and thermal fields, a flow field at a higher $\Ret$ is successively used for a one-step lower $\Ret$. Note that various statistical data and visualized fields are obtained after the scalar fields reached statistical-steady state. 
In the rest of the paper, all quantities with a superscript ($+$) are expressed in wall units. For instance, temperature is non-dimensionalized by the friction temperature. An overlined quantity, such as  , denotes an ensemble averaged component, and a primed one uf is a fluctuation component.

%%%%%%%%%%%%%%%%%%%%%%%%%%%%%%%%%
\begin{table}
  \begin{center}
\def~{\hphantom{0}}
  \begin{tabular}{cccccccc}
$\Ret$ & $\Rem$ & $\Rec$ & $L_x \times L_z$                & $N_x \times N_y \times N_z$  & $\Delta x^+$ & $\Delta y_{\rm min}^+$ & $\Delta z^+$\\
       &        &        &                                 &                              &              &                        & \\
 ~56   & 1590   & 1070   & $327.68\delta \times 128\delta$ & $4096 \times 96 \times 2048$ & 4.48         & 0.15                   & 3.50\\
 ~80   & 2330   & 1440   & $327.68\delta \times 128\delta$ & $4096 \times 96 \times 2048$ & 6.40         & 0.21                   & 5.00\\
 110   & 3270   & 1950   & $204.8\delta \times 64\delta$   & $4096 \times 96 \times 2048$ & 5.50         & 0.28                   & 3.44\\
 150   & 4640   & 2720   & $204.8\delta \times 64\delta$   & $4096 \times 96 \times 2048$ & 7.50         & 0.38                   & 4.69\\
  \end{tabular}
  \caption{Computational conditions with `huge' boxes: $L_i$, box size; $N_i$, number of grids; and $\Delta i$, spatial resolution. A quantity with the superscript ($+$) is non-dimensionalized by wall units.}
  \label{tab:hugebox}
  \end{center}
\end{table}

%%%%%%%%%%%%%%%%%%%%%%%%%%%%%%%%%
\begin{table}
  \begin{center}
\def~{\hphantom{0}}
  \begin{tabular}{cccccccc}
$\Ret$ & $\Rem$ & $\Rec$ & $L_x \times L_z$                & $N_x \times N_y \times N_z$  & $\Delta x^+$ & $\Delta y_{\rm min}^+$ & $\Delta z^+$\\
       &        &        &                                 &                              &              &                        & \\
 64 & 1850 & 1200 & $25.6\delta \times 12.8\delta$ & $512 \times 96 \times 256$ & 3.20 & 0.089 & 3.20\\
 70 & 2000 & 1260 & $25.6\delta \times 12.8\delta$ & $512 \times 96 \times 256$ & 3.50 & 0.097 & 3.50\\
 80--180 & 2290--5680 & 1400--3320 & $12.8\delta \times 6.4\delta$   & $512 \times 128 \times 256$ & 2.00--4.50 & 0.11--0.20 & 2.00--4.00\\
 \end{tabular}
  \caption{Computational conditions with `medium' boxes, which are of ordinary size as used in various studies on moderate Reynolds numbers. See Table 1 caption for legends.}
  \label{tab:middlebox}
  \end{center}
\end{table}

\section{Result and discussion}
In this section the visualization results for the turbulent puff in the channel flow are summarized first and then the statistics of turbulent heat transfer are discussed in detail.

\subsection{Visualization: Occurrence of turbulent puff}
\begin{figure}
 \begin{center}
  \textbf{a} \hspace{-1em}
  \includegraphics[width=80mm]{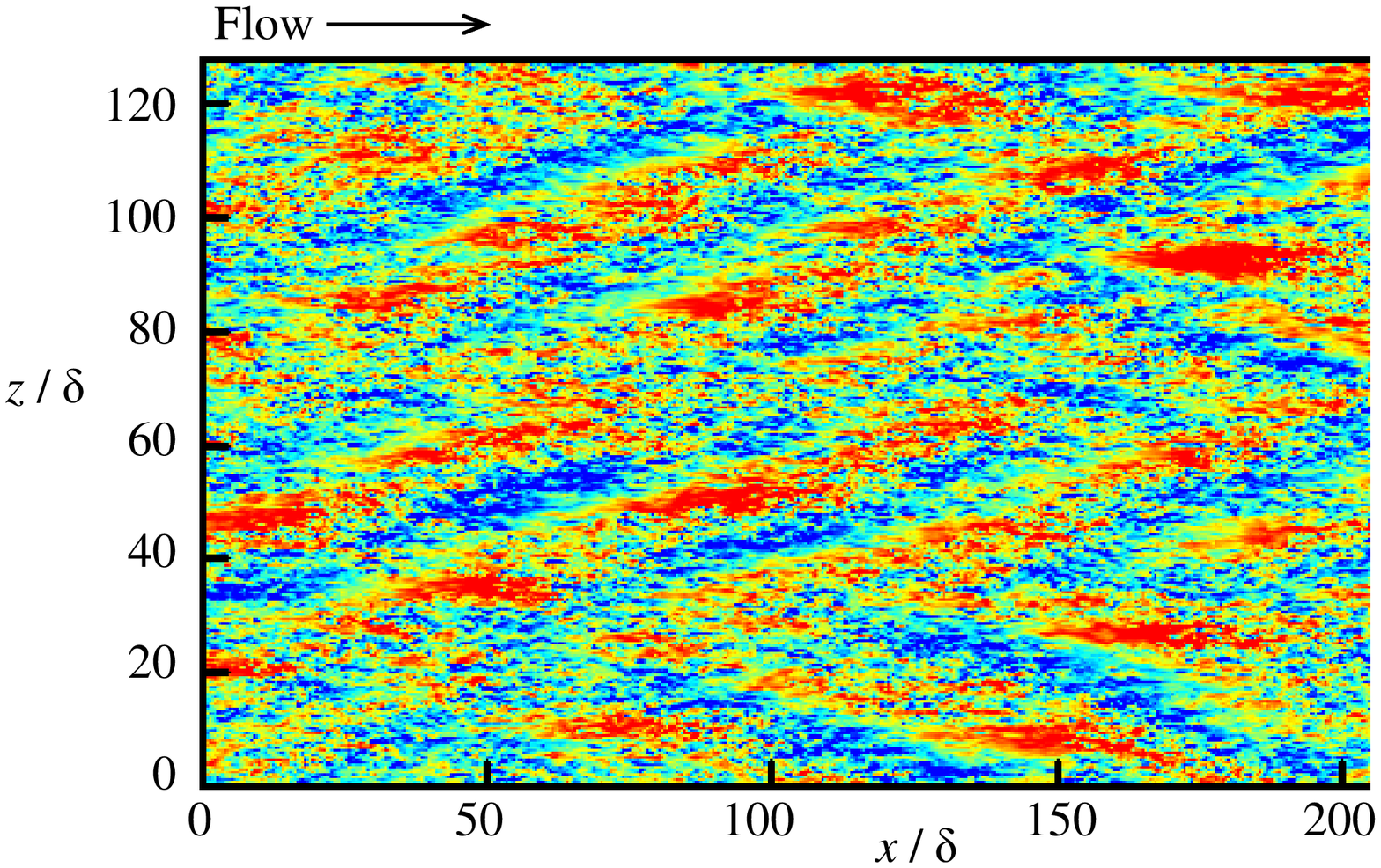}\\
  \textbf{b} \hspace{-1em}
  \includegraphics[width=80mm]{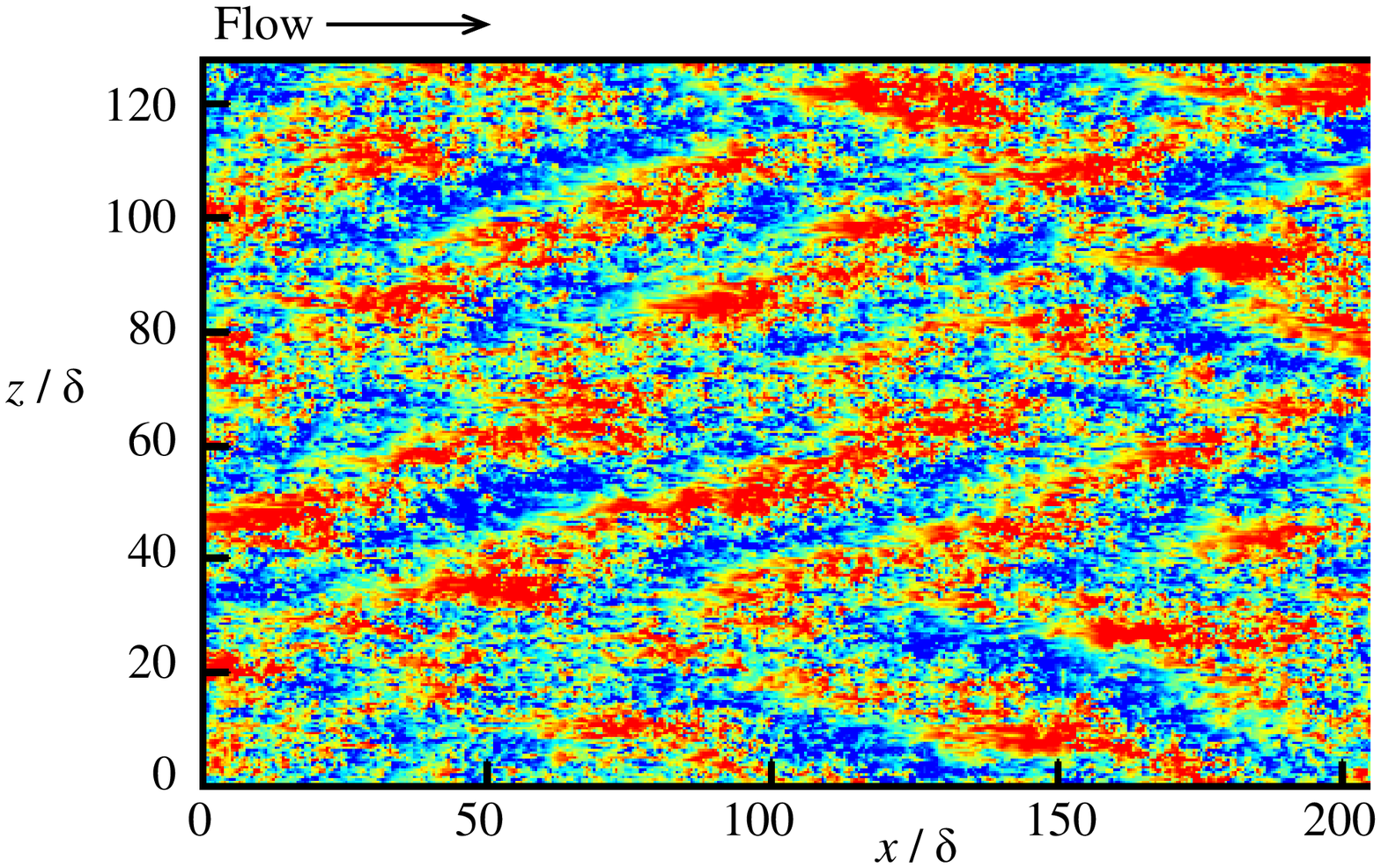}\\
  \textbf{c} \hspace{-1em}
  \includegraphics[width=80mm]{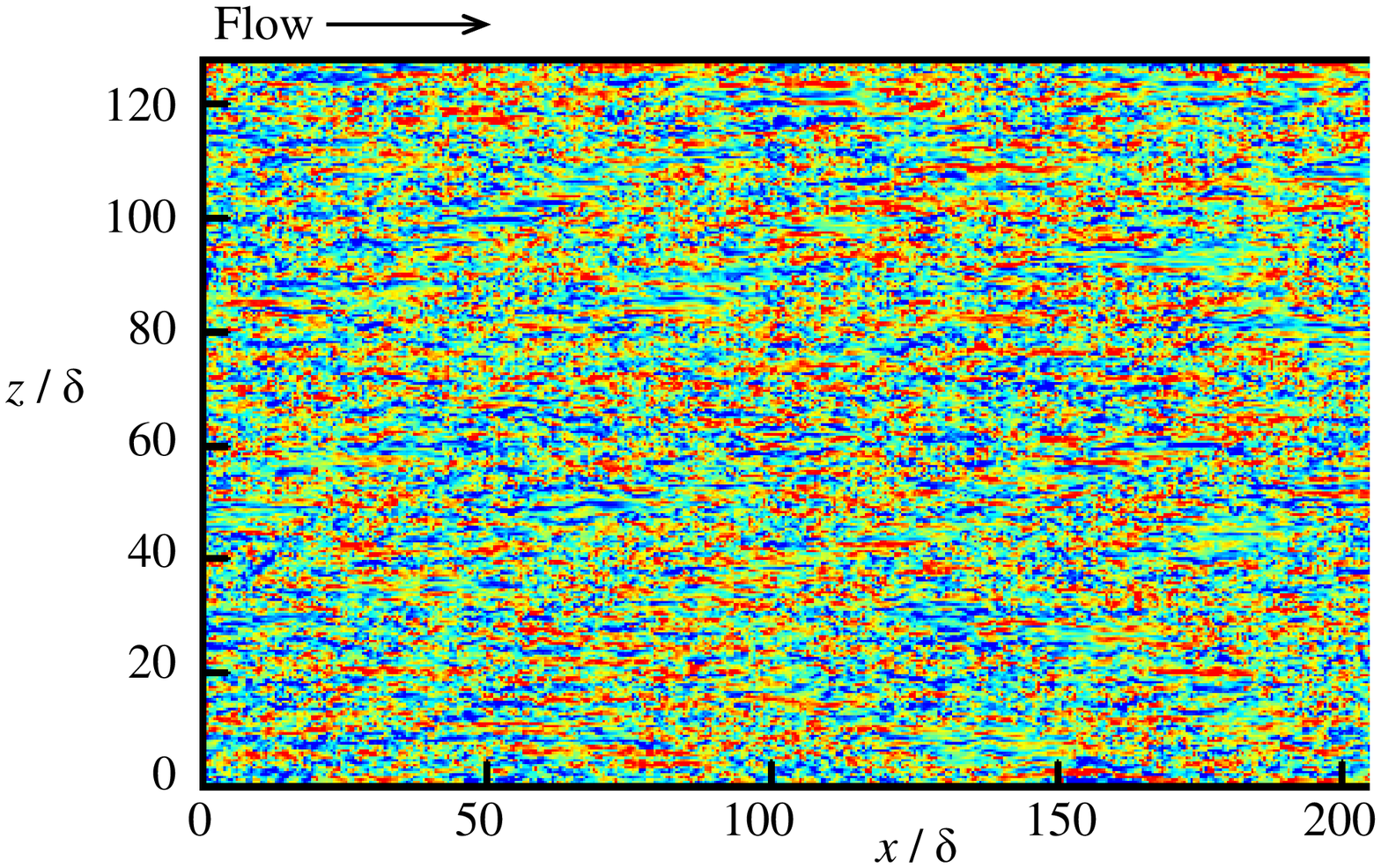}
  \caption{Contours of instantaneous flow and thermal fields in the ($x$, $z$) plane at the channel center for $\Ret = 80$ with the huge box: (a), $u'^+$; (b), $\theta'^+$ for UHF; and (c), $\theta'^+$ for CTD. Red-colored region represents positive quantities, i.e. (a) $u^+> 2.0$, (b) $\theta'^+ > 2.0$, and (c) $\theta'^+> 6.0$, and blue region represents negative quantities, i.e., (a) $u'^+< -2.0$, (b) $\theta'^+ < -2.0$, and (c) $\theta'^+< -6.0$. The field shown here is almost a half portion of the computational domain: namely, an area of ($L_x / 2 \times L_z$). }
  \label{fig:vis80}
 \end{center}
\end{figure}

\begin{figure}
 \begin{center}
  \textbf{a} \hspace{-1em}
  \includegraphics[width=80mm]{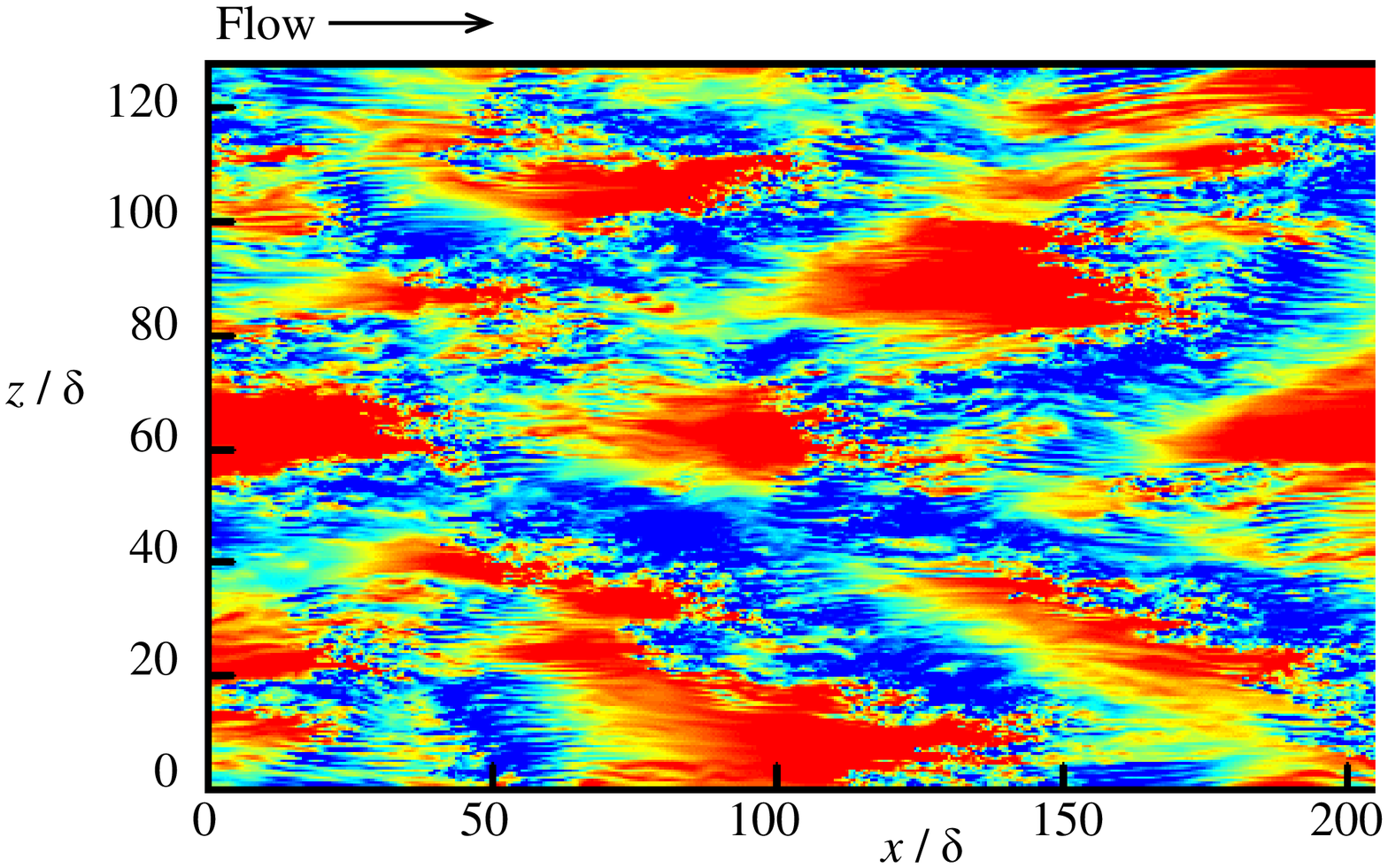}\\
  \textbf{b} \hspace{-1em}
  \includegraphics[width=80mm]{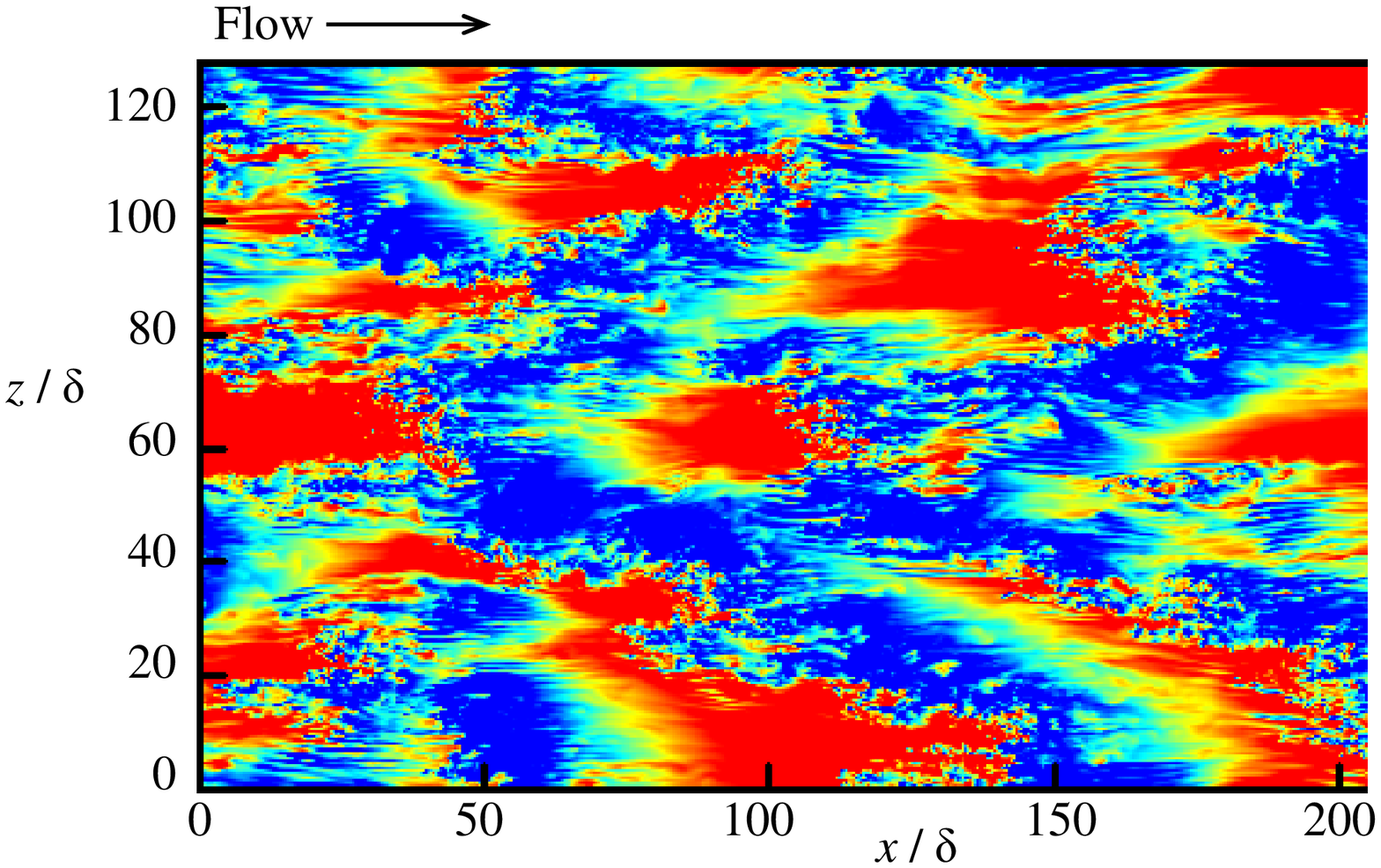}\\
  \textbf{c} \hspace{-1em}
  \includegraphics[width=80mm]{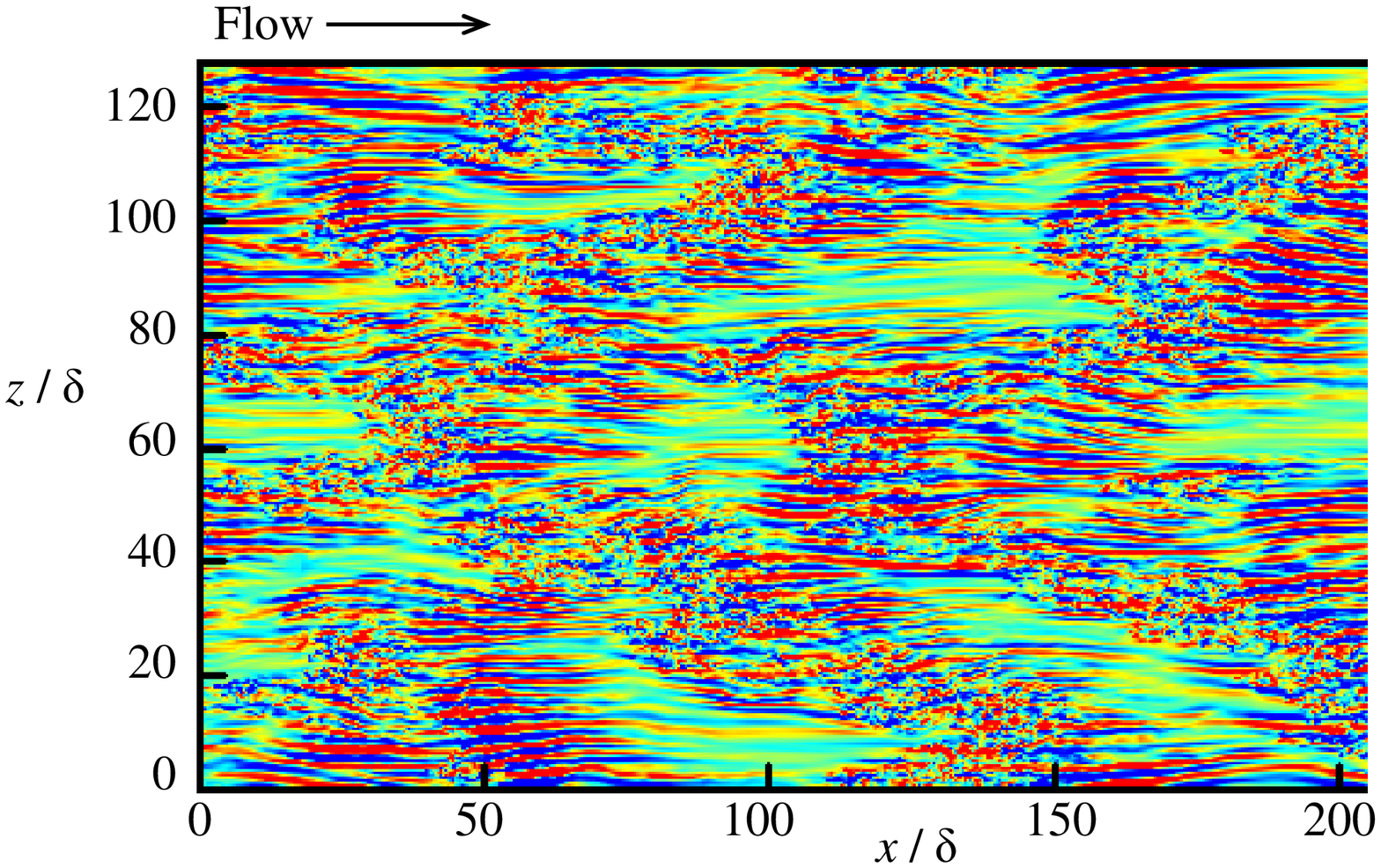}
  \caption{Same as \fref{fig:vis80} but for $\Ret=56$ with the huge box.}
  \label{fig:vis56}
 \end{center}
\end{figure}

We show a variation of the instantaneous field as a function of the Reynolds number. Figures \ref{fig:vis80}(a) and \ref{fig:vis56}(a) display snapshots of each flow with contours of the streamwise velocity fluctuation in the channel-central plane. The contours of the thermal fields for UHF and CTD are also shown in (b) and (c) of each figure, respectively. Both flow and thermal fields are let to stabilize before each snapshot is taken. For instance, a formation of the oblique stripes, which is observed in \fref{fig:vis80}(a), required a time more than $100\delta/u_\tau$ (about 8000 in wall units) to reach an equilibrium state. 

%%-------------------------------------- 

At $\Ret = 110$, the flow is homogeneously turbulent with respect to the horizontal directions, and both velocity and temperature fluctuations are rather randomly distributed (not shown here). The size of structures in the figure is about $7.0\delta \times 2.0\delta$ in $x$ and $z$ directions. This spanwise scale is slightly larger than that ($1.3\delta$--$1.6\delta$ given by \cite{Abe04a}) of the outer-layer structure at a high enough Reynolds number. When normalized in wall units, the scale of $770 \times 220$ is roughly comparable to that of the streaky structure in the near-wall region. This implies that the structures in the outer layer are affected by the near-wall structures due to a very low Reynolds number.

%%-------------------------------------- 

Decreasing $\Ret$ down to 80, there is a critical value below which a periodic structure appears with two preferred opposite inclinations. As you can see in \fref{fig:vis80}a, positive and negative fluctuations clearly exhibit a very large-scale pattern, which is tilted about 20$^\circ$--25$^\circ$ with respect to the streamwise direction. Based on the energy spectra of $u'$ (not shown here), the most energetic wavelengths are of the order of $65\delta$ and $26\delta$ in the stream- and spanwise directions, respectively: they correspond to the interval of the oblique stripes. The pattern moves at a constant velocity, which is almost same with the bulk mean velocity. A similar pattern has been observed in several configurations: e.g., gturbulent stripeh in a plane Couette flow \citep{Prigent03,Barkley07}, and gspiral turbulenceh in a Taylor-Couette flow \citep{Prigent03,Andereck86}. The present results indicate that features (tilted angle, interval and propagation speed) of the pattern would be more or less unchanged between Poiseuille flow and (plane/Taylor-) Couette flow. If you focus on an interface between an upstream high-speed region ($u' > 0$) and a downstream low-speed one ($u' < 0$), a highly disordered turbulent region appears. This spatial-intermittent turbulence is very similar to the puff in a pipe, as mentioned in the introduction.

\begin{figure}
 \begin{center}
  \textbf{a} \hspace{-1em}
  \includegraphics[width=80mm]{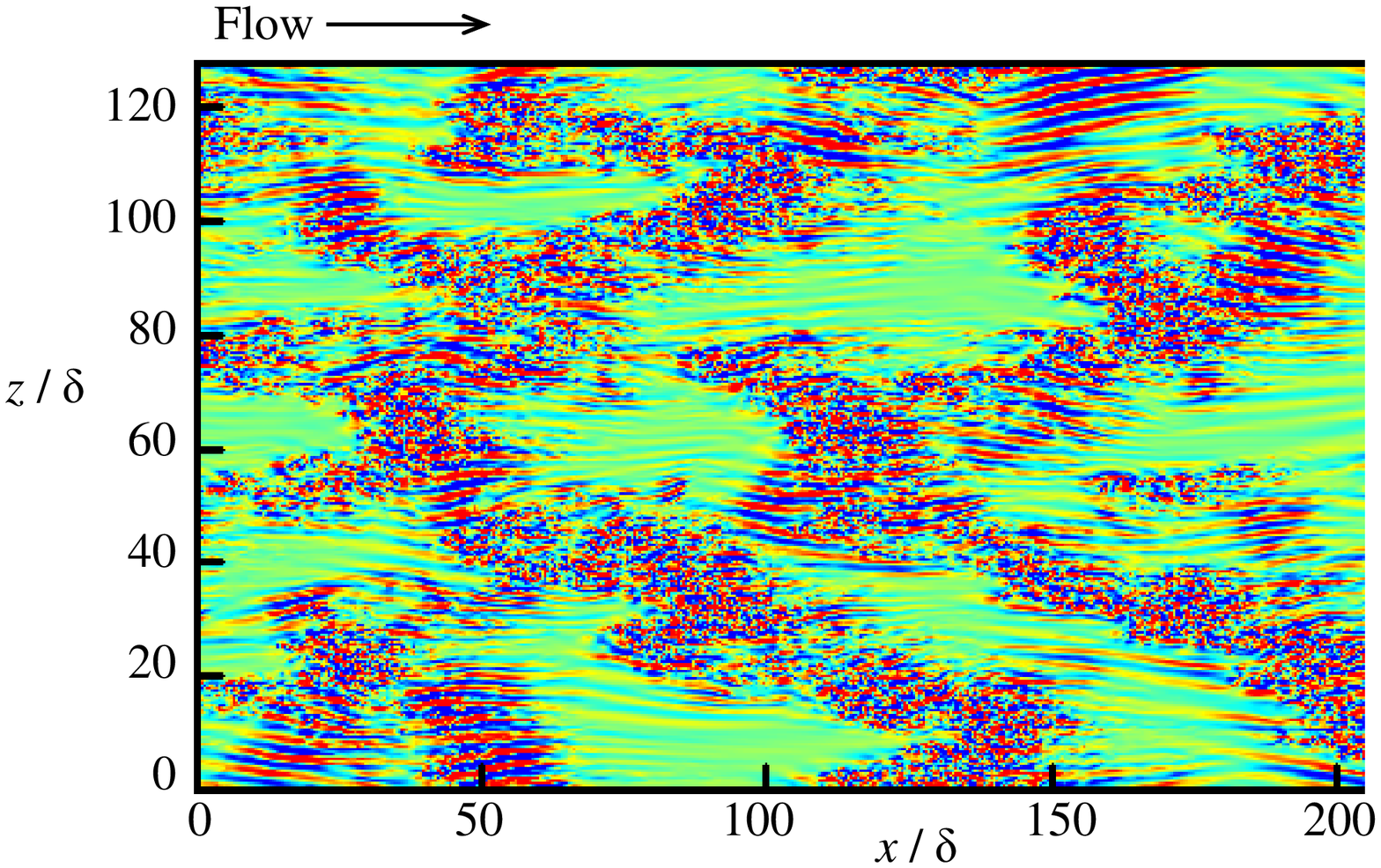}\\
  \textbf{b} \hspace{-1em}
  \includegraphics[width=80mm]{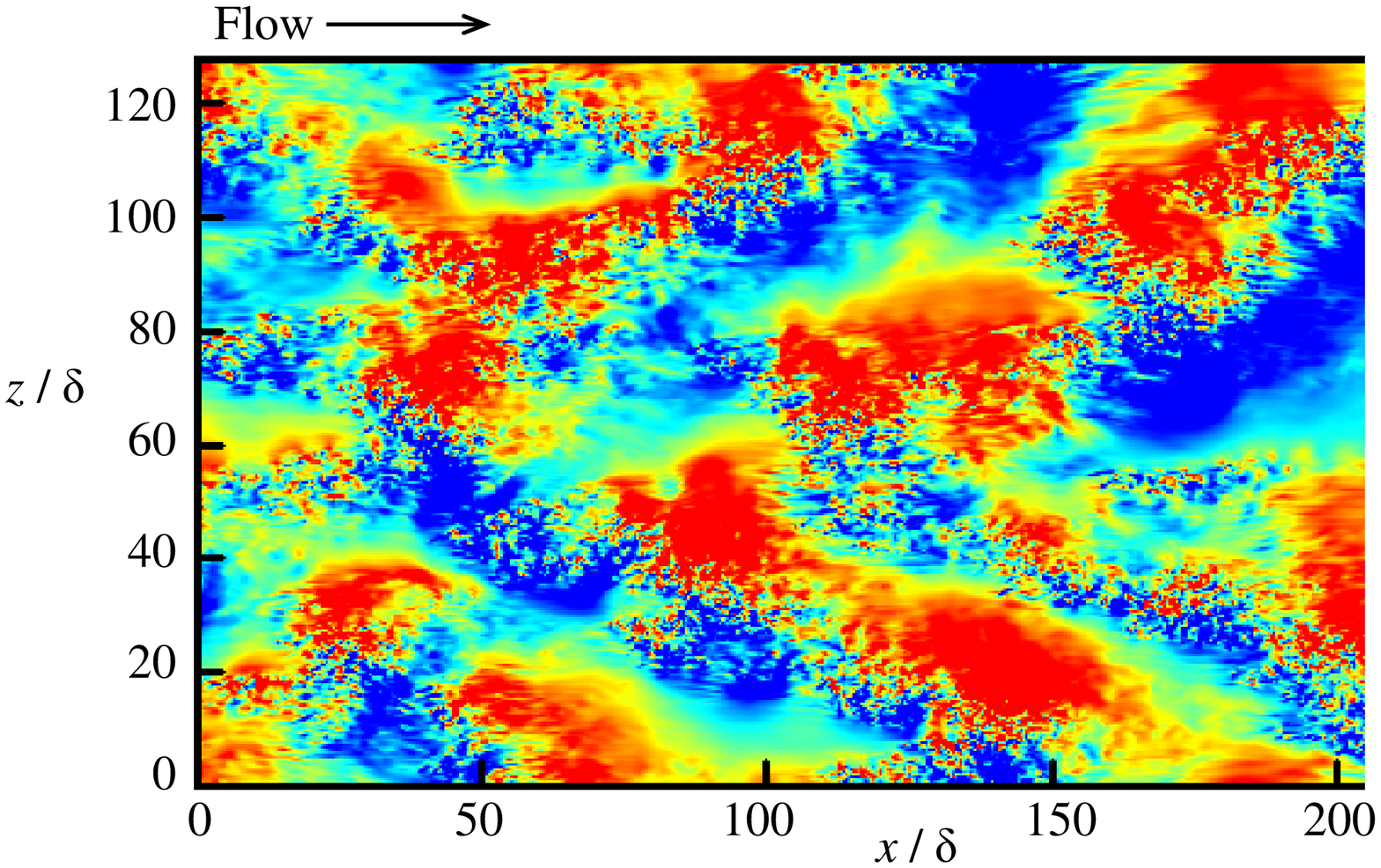}
  \caption{Contours of $v'$ and $w'$ at the same time and the same place with \fref{fig:vis56} for $\Ret=56$ with the huge box: red, (a) $v'^+> 0.5$, (b) $w'^+> 1.0$; blue, (a) $v'^+< -0.5$, (b) $w'^+< -1.0$.}
  \label{fig:vis56vw}
 \end{center}
\end{figure}

For even lower Reynolds number as low as $\Ret = 56$ ($\Rec = 1070$, $\Rem = 1590$), the stripes pattern breaks down, leaving a spatiotemporally intermittent turbulent regime, as given in \fref{fig:vis56}. This is consistent with the experimental results \citep{Patel69,Carlson82}. In \fref{fig:vis56vw}, the wall-normal velocity fluctuation (a) and the spanwise one (b) are shown in the same ($x$, $z$) plane and at the same time with the instantaneous field given in \fref{fig:vis56}. It can be seen from \fref{fig:vis56vw}(a) that the laminar and turbulent regions are clearly distinguished. Around the turbulent part, the much smoother and sparser contours of the disturbed laminar flow can be seen (almost green contours). In front of the turbulent region, a disturbed but non-turbulent region can be observed with elongated streaky structures. This streaky structure seems to correspond to the wave observed at the wingtip of the spot \citep{Carlson82,Henningson91}, not to the near-wall streaky structure. In contrast to the intermittent vf, the wf shows the large-scale structure, and is correlated with the $u'$. It is clear from figures \ref{fig:vis56}(a) and \ref{fig:vis56vw} that a high-speed fluid at upstream side of a turbulent region is tend to move along the inclination of the oblique pattern. 

%%-------------------------------------- 

As for the thermal field, a $\theta'$-distribution of UHF is almost identical to that of $u'$ due to the similarity of the boundary conditions (see figures \ref{fig:vis80}(b) and \ref{fig:vis56}(b)). On the other hand, the case of CTD shows that no large-scale organized structure was apparent at $\Ret = 80$ (see \fref{fig:vis80}(c)). When the Reynolds number is decreased down to $\Ret = 56$, a distinct influence of the puff can be observed, as seen in \fref{fig:vis56}(c). The intermittency of $\theta'$ is even more pronounced, with some streaks occurring in front of the turbulent region as similar to $v'$. This indicates that the wall-normal heat flux is intensified in the turbulent region and also in the region of the streaks.

\subsection{Turbulence statistics}

%%-------------------------------------- 
\begin{figure}
 \begin{center}
  \textbf{a} \hspace{-1em}
  \includegraphics[width=80mm]{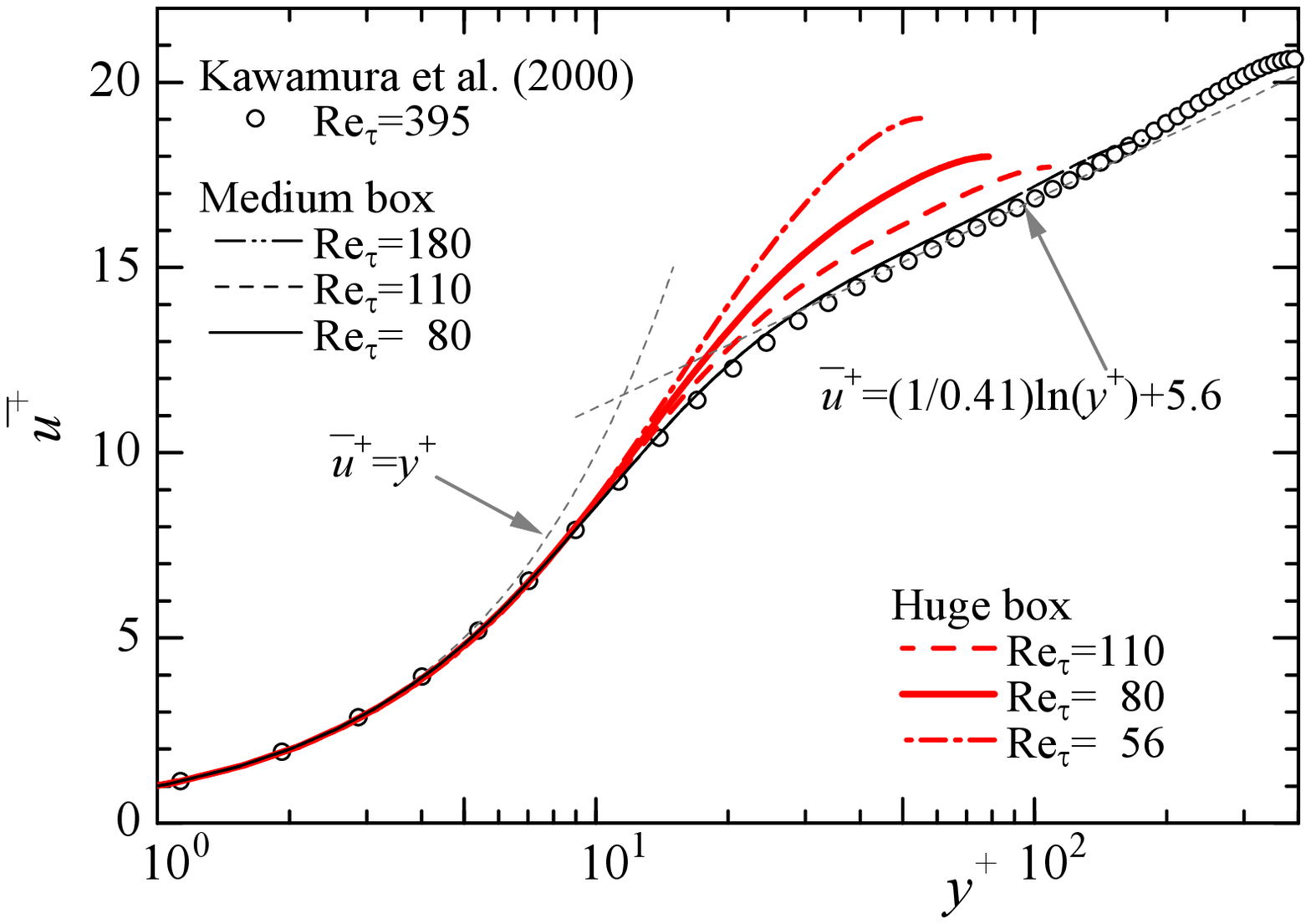} \\
  \textbf{b} \hspace{-1em}
  \includegraphics[width=80mm]{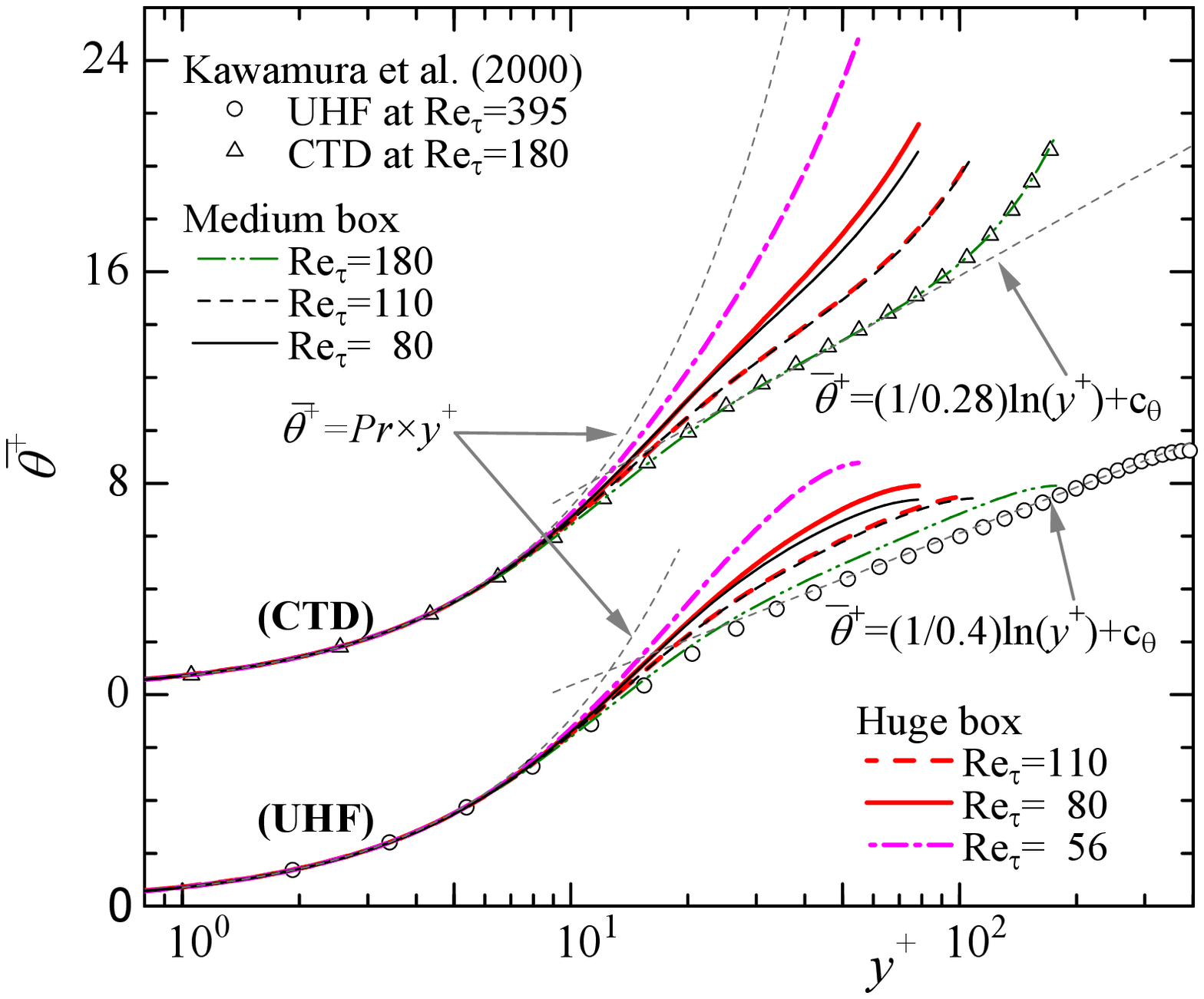}
  \caption{Mean velocity (a) and temperature (b) profiles in wall units..}
  \label{fig:mean}
 \end{center}
\end{figure}

The mean velocity and temperature profiles are presented as a function of $y^+$ in \fref{fig:mean}. In the present Reynolds number, the mean velocity distributions do not indicate an evident logarithmic region. The profiles deviate upward from ones of a high Reynolds number flow by \cite{Kawamura00}, and tend to be close to the laminar profile. Comparing profiles of different box sizes, the discrepancy for $\Ret = 110$ is minor throughout the channel, in contrast to the situation for $\Ret = 80$, where the values at the channel center remarkably increase with the expanding box size. This discrepancy is mainly because the puff appeared at $\Ret = 80$ as mentioned above. In addition, its large-scale pattern was not able to be captured in the case of the medium box (figure not shown here). Therefore, the channel-centerline value in the case of the huge box (i.e., with puff) is larger than that of the medium box (without puff), since the quasi-laminar region locally appeared in the flow field.

%%-------------------------------------- 
\begin{figure}
 \begin{center}
  \centerline{\includegraphics[width=80mm]{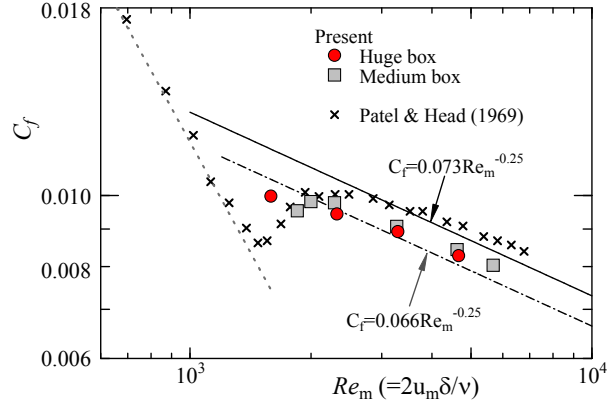}}
  \caption{Friction coefficient as a function of the bulk Reynolds number. The experimental result by \cite{Patel69} is also shown for comparison. Solid line is the empirical correlation function by \cite{Dean78}, chain line is the Blasius's friction law.}
  \label{fig:Cf}
 \end{center}
\end{figure}

%%-------------------------------------- 
\begin{figure}
 \begin{center}
  \centerline{\includegraphics[width=80mm]{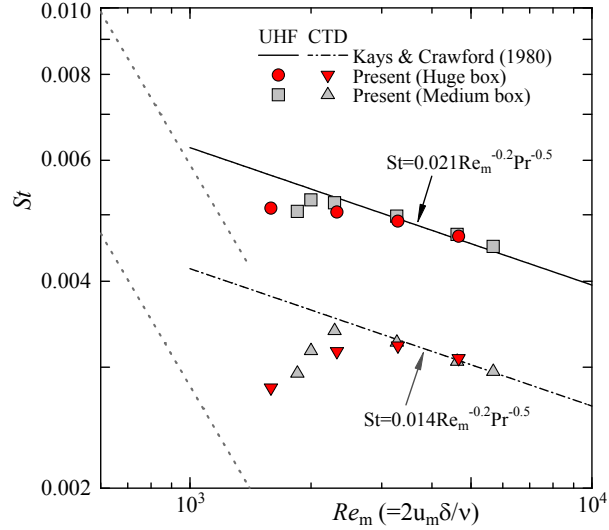}}
  \caption{Stanton number for each thermal condition {\it vs} $\Rem$.}
  \label{fig:St}
 \end{center}
\end{figure}

Figures \ref{fig:Cf} and \ref{fig:St} show variations of the skin friction coefficient ($C_f$) and the Stanton number (\textit{St}) in comparison with the empirical correlations for a fully turbulent regime \citep{Dean78,Kays80}. They are non-dimensionalized as follows:
\begin{eqnarray}
C_f & = & \frac{2\tau_{\rm wall}}{\rho u_m^+} = \frac{2}{{u_m^+}^2} \\
\textit{St} & = & \frac{\textit{Nu}}{\Rem \Prn} = \frac{2}{u_m^+ \theta_m^+}
\end{eqnarray}
where $\tau_{\rm wall}$, $\rho$, and $\theta_{\rm m}$ are the averaged wall shear stress, the density and the bulk mean temperature, respectively. No significant dependence on the box size is found in the fully turbulent regime of $\Rem > 3000$. Especially, the obtained Stanton number is good agreement with the function proposed by \cite{Kays80}. In the transitional regime of $\Rem < 3000$, both $C_f$ and $\textit{St}$ with the huge box, accompanied by the puff, tend to show a gradual variation compared to the medium box (without puff); and they stay closer to the empirical correlations even for very low Reynolds numbers ($\Rem < 2000$). This clearly illustrates that the puff should be effective in sustaining turbulence and in heat transfer enhancement.

\section{Summary and conclusions}
\label{sec:concl}

With a large computational domain of $327.68\delta \times 2\delta \times 128\delta$, we demonstrate that the puff in the transitional channel flow is not an artifact of the limited lengths of the computational domain. For the transitional Reynolds numbers range of about $\Rem = 1500$--3000, the puff sustains turbulence in localized region, where the heat transfer is enhanced. 

\section*{Acknowledgments}
The present computations were performed with the use of the supercomputing resources at Cyberscience Center of Tohoku University. The first author (T.T.) was supported by Japan Society for the Promotion Science Fellowship (18-81). This work was conducted in Research Center for the Holistic Computational Science (Holcs) supported by MEXT.

\vspace{1em}

\noindent
This paper is a revised and expanded version of a paper entitled ``Turbulent heat transfer in a channel flow at transitional Reynolds numbers'', presented by T. Tsukahara and H. Kawamura, at 1st Asian Symposium on Computational Heat Transfer and Fluid Flow, 18-21  October 2007, Xi'an, China.

\bibliographystyle{jfm}

\bibliography{tsuka}

\begin{thebibliography}{20}
\expandafter\ifx\csname natexlab\endcsname\relax\def\natexlab#1{#1}\fi

\bibitem[Abe {\em et~al.\/}(2004{\natexlab{{\em a\/}}})Abe, Kawamura \&
  Choi]{Abe04a}
{\sc Abe, H., Kawamura, H. \& Choi, H.} 2004{\natexlab{{\em a\/}}} Very
  large-scale structures and their effects on the wall shear-stress
  fluctuations in a turbulent channel flow up to $\textit{Re}_\tau =640$. {\em
  J. Fluids Eng.\/} {\bf 126}, 835--843.

\bibitem[Abe {\em et~al.\/}(2004{\natexlab{{\em b\/}}})Abe, Kawamura \&
  Matsuo]{Abe04b}
{\sc Abe, H., Kawamura, H. \& Matsuo, Y.} 2004{\natexlab{{\em b\/}}} Surface
  heat-flux fluctuations in a turbulent channel flow up to $\textit{Re}_\tau =
  1020$ with $\textit{Pr} = 0.025$ and 0.71. {\em Int. J. Heat and Fluid
  Flow\/} {\bf 25}, 404--419.

\bibitem[Andereck {\em et~al.\/}(1986)Andereck, Liu \& Swinney]{Andereck86}
{\sc Andereck, C.~D., Liu, S.~S. \& Swinney, H.~L.} 1986 Flow regimes in a
  circular {C}ouette system with independently rotating cylinders. {\em
  J.~Fluid Mech.\/} {\bf 164}, 155--183.

\bibitem[Barkley \& Tuckerman(2007)]{Barkley07}
{\sc Barkley, D. \& Tuckerman, L.~S.} 2007 Mean flow of turbulent-laminar
  patterns in plane {C}ouette flow. {\em J.~Fluid Mech.\/} {\bf 576}, 109--137.

\bibitem[Carlson {\em et~al.\/}(1982)Carlson, Widnall \& Peeters]{Carlson82}
{\sc Carlson, D.~R., Widnall, S.~E. \& Peeters, M.~F.} 1982 A
  flow-visualization study of transition in plane {P}oiseuille flow. {\em J.
  Fluid Mech.\/} {\bf 121}, 487--505.

\bibitem[Dean(1978)]{Dean78}
{\sc Dean, R.~D.} 1978 Reynolds number dependence of skin friction and other
  bulk flow variables in two-dimensional rectangular duct flow. {\em J. Fluids
  Eng.\/} {\bf 100}, 215--222.

\bibitem[Henningson \& Kim(1991)]{Henningson91}
{\sc Henningson, D.~S. \& Kim, J.} 1991 On turbulent spots in plane poiseuille
  flow. {\em J. Fluid Mech.\/} {\bf 228}, 183--205.

\bibitem[Kasagi {\em et~al.\/}(1992)Kasagi, Tomita \& Kuroda]{Kasagi92}
{\sc Kasagi, N., Tomita, Y. \& Kuroda, A.} 1992 A direct numerical simulation
  for passive scalar field in a turbulent channel flows. {\em J. Heat
  Transfer\/} {\bf 114}, 598--606.

\bibitem[Kawamura {\em et~al.\/}(2000)Kawamura, Abe \& Shingai]{Kawamura00}
{\sc Kawamura, H., Abe, H. \& Shingai, K.} 2000 {DNS} of turbulence and heat
  transport in a channel flow with different reynolds and prandtl numbers and
  boundary conditions. In {\em Proceedings of Third International Symposium on
  Turbulence, Heat and Mass Transfer\/} (ed. Y.~Nagano {\em et~al.\/}), pp.
  15--32. Nagoya, Japan.

\bibitem[Kawamura {\em et~al.\/}(1998)Kawamura, Ohsaka, Abe \&
  Yamamoto]{Kawamura98}
{\sc Kawamura, H., Ohsaka, K., Abe, H. \& Yamamoto, K.} 1998 {DNS} of turbulent
  heat transfer in channel flow with low to medium-high prandtl number fluid.
  {\em Int. J. Heat and Fluid Flow\/} {\bf 19}, 482--491.

\bibitem[Kays \& Crawford(1980)]{Kays80}
{\sc Kays, W.~M. \& Crawford, M.~E.} 1980 {\em Convective heat and mass
  transfer, 2nd ed.\/}. McGraw-Hill, New York.

\bibitem[Kim \& Moin(1989)]{Kim89}
{\sc Kim, J. \& Moin, P.} 1989 Transport of passive scalars in a turbulent
  channel flow. In {\em Turbulent Shear Flows 6\/} (ed. J.-C.~Andr\'e {\em
  et~al.\/}), pp. 85--96.

\bibitem[Lyons {\em et~al.\/}(1991)Lyons, Hanratty \& Mclaughlin]{Lyons91}
{\sc Lyons, S.~L., Hanratty, T.~J. \& Mclaughlin, J.~B.} 1991 Direct numerical
  simulation of passive scalar heat transfer in a turbulent channel flow. {\em
  Int. J. Heat and Mass Transfer\/} {\bf 39}, 1149--1161.

\bibitem[Na \& Hanratty(2000)]{Na00}
{\sc Na, Y. \& Hanratty, T.~J.} 2000 Limiting behavior of turbulent scalar
  transport close to a wall. {\em Int. J. Heat and Mass Transfer\/} {\bf 43},
  1749--1758.

\bibitem[Patel \& Head(1969)]{Patel69}
{\sc Patel, V.~C. \& Head, M.~R.} 1969 Some observations on skin friction and
  velocity profiles in fully developed pipe and channel flows. {\em J. Fluid
  Mech.\/} {\bf 38}, 181--201.

\bibitem[Prigent {\em et~al.\/}(2003)Prigent, Gr\'egoire, Chat\'e \&
  Dauchot]{Prigent03}
{\sc Prigent, A., Gr\'egoire, G., Chat\'e, H. \& Dauchot, O.} 2003
  Long-wavelength modulation of turbulent shear flows. {\em Physica D\/} {\bf
  174}, 100--113.

\bibitem[Tsukahara {\em et~al.\/}(2006)Tsukahara, Iwamoto, Kawamura \&
  Takeda]{Tsukahara2006}
{\sc Tsukahara, T., Iwamoto, K., Kawamura, H. \& Takeda, T.} 2006 {DNS} of heat
  transfer in a transitional channel flow accompanied by a turbulent puff-like
  structure. In {\em Proceedings of Fifth International Symposium on
  Turbulence, Heat and Mass Transfer\/} (ed. K.~Hanjali\'c {\em et~al.\/}), pp.
  193--196. Dubrovnik, Croatia.

\bibitem[Tsukahara {\em et~al.\/}(2005)Tsukahara, Seki, Kawamura \&
  Tochio]{Tsukahara2005}
{\sc Tsukahara, T., Seki, Y., Kawamura, H. \& Tochio, D.} 2005 {DNS} of
  turbulent channel flow at very low {R}eynolds numbers. In {\em Proceedings of
  Fourth International Symposium on Turbulence and Shear Flow Phenomena\/} (ed.
  J.~A. C.~Humphrey {\em et~al.\/}), pp. 935--940. Williamsburg, USA.

\bibitem[Wygnanski {\em et~al.\/}(1975)Wygnanski, Sokolov \&
  Friedman]{Wygnanski75}
{\sc Wygnanski, I., Sokolov, M. \& Friedman, D.} 1975 On transition in a pipe.
  {P}art 2. {T}he equilibrium puff. {\em Journal of Fluid Mechanics\/} {\bf
  69}, 283--304.

\bibitem[Wygnanski \& Champagne(1973)]{Wygnanski73}
{\sc Wygnanski, I.~J. \& Champagne, F.~H.} 1973 On transition in a pipe. {P}art
  1. {T}he origin of puffs and slugs and the flow in a turbulent slug. {\em J.
  Fluid Mech.\/} {\bf 59}, 281--335.

\end{thebibliography}

\end{document}